\documentclass[%
 byrevtex,
 reprint,
amsmath,amssymb,
 aps,
 prb,
longbibliography,
 lengthcheck,%
]{revtex4-1}

\providecommand{\hm}{\pmb\bm}
\providecommand{\mbf}{\mathbf}

\usepackage{microtype}
\usepackage[latin1]{inputenc}
\usepackage{mathptmx}
\usepackage{times}
\usepackage{bm}
\usepackage{mathtools}
\usepackage{graphicx}
\usepackage{xcolor}
\usepackage{dcolumn}
\usepackage{comment}

\RequirePackage{xspace}

\newcommand*{\eg}{\emph{e.g.}\xspace}

\newcommand*{\ie}{\emph{i.e.}\xspace}

\newcommand*{\eqn}[1]{Eq.~(\ref{#1})}
\newcommand*{\eqns}[1]{Eqs.~(\ref{#1})}

\newcommand*{\upd}[1]{\mathrm{d}#1}
\newcommand*{\ddt}[1]{\frac{\upd{#1}}{\upd{t}}}

\newcommand*{\dS}{\upd{^2}\!x}
\newcommand*{\dSvec}[1]{\dS{#1}\,\nunit{#1}}
\newcommand*{\ointSvec}[2]{\oint_{S{#1}}\!\dSvec{#1}\bdot{#2}}

\newcommand*{\Ave}[1]{\mathinner{\left\langle{#1}\right\rangle}}
\newcommand*{\cc}[1]{{#1}^*}
\newcommand*{\eps}{\varepsilon}
\newcommand*{\epsz}{\eps_{0}}

\newcommand*{\im}{{\mathrm{i}}}

\newcommand*{\Mag}[1]{\mathinner{\left|{#1}\right|}}

\renewcommand*{\Re}[1]{\mathrm{Re}\left\{#1\right\}}
\newcommand*{\sub}[1]{_{\mathrm{#1}}}
\newcommand*{\Sup}[1]{^{\mathrm{#1}}}
\newcommand*{\tret}{t\sub{ret}}
\newcommand*{\w}{\omega}

\renewcommand*{\vec}[1]{\mbf{#1}}
\newcommand*{\0}{\vec{0}}
\newcommand*{\B}{\vec{B}}
\newcommand*{\E}{\vec{E}}
\newcommand*{\g}{\vec{g}}
\newcommand*{\gfield}{\g\Sup{field}}
\newcommand*{\h}{\vec{h}}
\newcommand*{\hfield}{\h\Sup{field}}
\newcommand*{\I}{\vec{I}}
\newcommand*{\Idot}{\Dot{\I}}
\newcommand*{\J}{\vec{J}}
\newcommand*{\Jfield}{\J\Sup{field}}
\newcommand*{\Jmech}{\J\Sup{mech}}
\renewcommand*{\j}{\vec{j}}
\newcommand*{\jdot}{\Dot{\j}}

\renewcommand*{\k}{\vec{k}}

\renewcommand*{\L}{\vec{L}}
\newcommand*{\Lfield}{\L\Sup{field}}
\newcommand*{\unit}[1]{\hat{\mbf{#1}}}
\newcommand*{\kunit}{\unit{k}}
\newcommand*{\nunit}{\unit{n}}
\newcommand*{\p}{\vec{p}}
\newcommand*{\pfield}{\p\Sup{field}}
\newcommand*{\pmech}{\p\Sup{mech}}
\renewcommand*{\S}{\vec{S}}
\newcommand*{\Sfield}{\bm\Sigma\Sup{field}}
\newcommand*{\x}{\vec{x}}
\newcommand*{\xx}{\x-\x}

\newcommand*{\ten}[1]{\pmb{\mathsf{#1}}}

\newcommand*{\grad}{\hm{\nabla}}
\newcommand*{\bdot}{\bm\cdot}

\newcommand*{\cross}{\bm\times}

\newcommand*{\dV}{\upd{^3}\!x}
\newcommand*{\intV}[2]{\int_{V{#1}}\!\dV{#1}\,{#2}}	

\begin{document}


\title{Linear and angular momentum of electromagnetic fields generated
 by an arbitrary distribution of charge and current densities at rest}

\author{B.\,Thid\'e}
\affiliation{%
 Swedish Institute of Space Physics,
 {\AA}ngstr\"{o}m Laboratory,
 P.\,O.~Box~521,
 SE-751\,21,
 Uppsala,
 Sweden}%

\author{H.\,Then}
 \affiliation{%
 Institute of Physics, 
 Carl-von-Ossietzky Universit\"at Oldenburg,
 D-261\,11 Oldenburg,
 Germany}%

\author{F.\,Tamburini}
 \affiliation{%
 Department of Astronomy,
 University of Padova,
 vicolo dell' Osservatorio 3,
 Padova,
 Italy}%

\author{J.\,Lindberg}
\affiliation{%
 Uppsala University,
 Uppsala,
 Sweden}

\revised{\today}

\begin{abstract}

Starting from Stratton-Panofsky-Phillips-Jefimenko equations for
the electric and magnetic fields generated by completely arbitrary
charge and current density distributions at rest, we derive far-zone
approximations for the fields, containing all components, dominant
as well as sub-dominant. Using these approximate formulas, we derive
general formulas for the total electromagnetic linear momentum and
angular momentum, valid at large distances from arbitrary, non-moving
charge and current sources.

\end{abstract}

\pacs{PACS here\ldots}

\maketitle

\section{Introduction}

Standard classical electrodynamics offers different methods for
calculating electromagnetic fields and physical observables derived from
them. Usually the methods amount to solving the Maxwell equations in
one form or another, or to find the potentials, choose a gauge and then
perform the calculations.

In recent years, the use of alternative methods, based on integral
equations that relate the fields directly to their charge and current
sources, have gained attraction; see Ref.~\onlinecite{Souza&al:AJP:2009}
and references cited therein. Here we show how these methods can be
used to directly derive general expressions for the linear and angular
momentum of the electromagnetic field that approximate these quantities
at large distances from the sources. Among other things,
these expressions show that both linear momentum and angular momentum
can be used to transfer information wirelessly in free space.

In Chapter VIII of his 1941 textbook, \textcite{Stratton:Book:1941}
calculates, without the intervention of potentials, temporal
Fourier transform expressions for the retarded electric and
magnetic fields, $\E(t,\x)$ and $\B(t,\x)$, respectively, generated
by arbitrary distributions of charge and current densities at
rest relative to the observer. In Chapter~14 of the second
edition of their textbook on electrodynamics, published in 1962,
\textcite{Panofsky&Phillips:Book:1962} present a variant form
of these expressions, and give them also in ordinary space-time
coordinates. Four years later, Jefimenko published his electrodynamics
textbook\cite{Jefimenko:Book:1966} where the Panofsky and Phillips
expressions for the retarded $\E$ and $\B$ fields were given in
Chapter~15. These expressions are sometimes referred to as the Jefimenko
equations,\cite{Heald&Marion:Book:1995,Jackson:Book:1999} but should,
perhaps, rather be called the Stratton-Panofsky-Phillips-Jefimenko
(SPPJ) equations, a name we will use throughout.

Introducing the notation $\x$ for the observer's coordinate, $\x'$ for
the source coordinate, and
\begin{align}
\label{eq:tret}
\tret'(t,\x,\x') = t - \frac{\Mag{\xx'}}{c}
\end{align}
for the retarded time\footnote{We do not consider advanced time
solutions.} relative to the source point where $t$ is the observer's
time and $c$ is the speed of light, the SPPJ equations for the retarded
electric and magnetic fields can, in obvious notation, be written
\begin{subequations}
\label{eq:EB_Jefimenko}
\begin{align}
\label{eq:E_Jefimenko}
\E(t,\x) ={}& \frac{1}{4\pi\epsz}
 \intV{'}{\frac{\rho(\tret',\x')}{\Mag{\xx'}^2}}\,\frac{\xx'}{\Mag{\xx'}}
\notag\\
 &+\frac{1}{4\pi\epsz c}\intV{'}{\frac{\dot\rho(\tret',\x')}{\Mag{\xx'}}}
  \,\frac{\xx'}{\Mag{\xx'}}
\notag\\
 &-\frac{1}{4\pi\epsz c^2}\intV{'}{\frac{\dot\j(\tret',\x')}{\Mag{\xx'}}}
\intertext{and}
\label{eq:B_Jefimenko}
\B(t,\x) ={}& \frac{1}{4\pi\epsz c^2}
   \intV{'}{\frac{\j(\tret',\x')\cross(\xx')}{\Mag{\xx'}^3}}
\notag\\
&+\frac{1}{4\pi\epsz c^3}
   \intV{'}{\frac{\jdot(\tret',\x')\cross(\xx')}{\Mag{\xx'}^2}}
\end{align}
\end{subequations}
respectively.  By introducing the vector
\begin{multline}
\vec{T}(\tret',\x',\x) = \frac{\rho(\tret',\x')}{\Mag{\xx'}^2}
 \,\frac{\xx'}{\Mag{\xx'}}
\\
 +\frac{1}{c\Mag{\xx'}^2}\bigg(
 \j(\tret',\x')\bdot\frac{\xx'}{\Mag{\xx'}}\bigg)\frac{\xx'}{\Mag{\xx'}} 
\\
 +\frac{1}{c\Mag{\xx'}^2}
 \bigg(\j(\tret',\x')\cross\frac{\xx'}{\Mag{\xx'}}\bigg)
 \cross\frac{\xx'}{\Mag{\xx'}}
\\
 +\frac{1}{c^2\Mag{\xx'}}
 \bigg(\dot\j(\tret',\x')\cross\frac{\xx'}{\Mag{\xx'}}\bigg)
 \cross\frac{\xx'}{\Mag{\xx'}}
\end{multline}
the SPPJ equations (\ref{eq:EB_Jefimenko}) can be cast into the
more compact and symmetric form
\begin{subequations}
\begin{align}
\E(t,\x) ={}& \frac{1}{4\pi\epsz}\intV{'}{\vec{T}(\tret',\x',\x)} 
\\
\B(t,\x) = {}&\frac{1}{4\pi\epsz c}
 \intV{'}{\frac{\xx'}{\Mag{\xx'}}\cross\vec{T}(\tret',\x',\x)}
\end{align}
\end{subequations}
so that \eqn{eq:E_Jefimenko} can be written in the alternative form
\begin{multline}
\label{eq:E_Thide}
\E(t,\x) = 
 \frac{1}{4\pi\epsz}
 \intV{'}{\frac{\rho(\tret',\x')(\xx')}{\Mag{\xx'}^3}}
\\
+\frac{1}{4\pi\epsz c}
 \intV{'}{\frac{[\j(\tret',\x')\bdot(\xx')](\xx')}{\Mag{\xx'}^4}}
\\
+ \frac{1}{4\pi\epsz c}
 \intV{'}{\frac{[\j(\tret',\x')\cross(\xx')]\cross(\xx')}{\Mag{\xx'}^4}}
\\
+\frac{1}{4\pi\epsz c^2}
 \intV{'}{\frac{[\jdot(\tret',\x')\cross(\xx')]\cross(\xx')}{\Mag{\xx'}^3}}
\end{multline}
which more clearly exhibits the relation between the various components
of the retarded $\E$ field.

In Section II, far-zone formulas for the $\E$ and $\B$ fields, based on the
SPPJ equations, are introduced.  These equations are used in Section III
and IV to derive far-zone expressions for the electromagnetic linear momentum
and angular momentum, respectively. In Section V, a summary and conclusions
are made.

\section{The fields at large distances from the source}

\begin{figure}
\resizebox{0.95\columnwidth}{!}{%
 \includegraphics{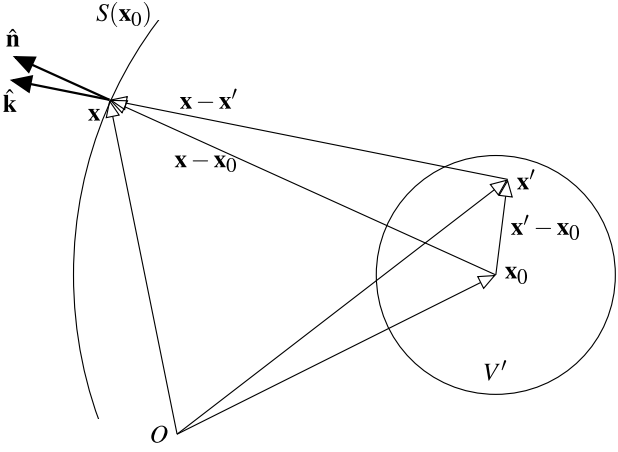}
}
\caption[Radiation in the far zone]{
The sources are located at points $\x'$ inside a volume $V'$ located
around $\x_0$ such that ${\sup\Mag{\x'-\x_0}\ll\inf\Mag{\xx'}}$, where $\x$
is the observation point. The unit vector $\kunit$ is directed along $\xx'$.
$\nunit$ is along $\xx_0$ and orthogonal to the surface $S(\x_0)$ with
its origin at $\x_0$.
}
\label{fig:farzone}
\end{figure}

Following \textcite{Panofsky&Phillips:Book:1962} we introduce the
temporal Fourier component representations of \eqn{eq:E_Thide} 
for $\E(t,\x)$ and of \eqn{eq:B_Jefimenko} for $\B(t,\x)$, 
\begin{subequations}
\label{eq:EB_w}
\begin{align}
\label{eq:E_w}
\E_\w(\x) ={}& \frac{1}{4\pi\epsz}\Big(
  \intV{'}{\frac{\rho_\w(\x')e^{\im k\Mag{\xx'}}(\xx')}{\Mag{\xx'}^3}}
\notag\\
 &+\frac{1}{c}\intV{'}
  {\frac{[\j_\w(\x')e^{\im k\Mag{\xx'}}\bdot(\xx')](\xx')}{\Mag{\xx'}^4}}
\notag\\
 &+\frac{1}{c}\intV{'}
  {\frac{[\j_\w(\x')e^{\im k\Mag{\xx'}}\cross(\xx')]\cross(\xx')}
  {\Mag{\xx'}^4}}
\notag\\
 &-\frac{\im k}{c}\intV{'}
  {\frac{[\j_\w(\x')e^{\im k\Mag{\xx'}}\cross(\xx')]\cross(\xx')}
  {\Mag{\xx'}^3}}\Big)
\end{align}
and
\begin{align}
\label{eq:B_w}
\B_\w(\x) ={}&
\frac{1}{4\pi\epsz c^2}\Big(\intV{'}{\frac{\j_\w(\x')e^{\im k\Mag{\xx'}}
 \cross(\xx')}{\Mag{\xx'}^3}}
\notag\\
&+\intV{'}{\frac{(-\im k)\j_\w(\x')e^{\im k\Mag{\xx'}}
 \cross(\xx')}{\Mag{\xx'}^2}}\Big)
\end{align}
\end{subequations}
respectively.

As illustrated in Fig.~\ref{fig:farzone}, the observation point
$\x$ is assumed to be located far away from the sources,
which in turn are assumed to be localized near a point $\x_0$
inside a volume $V'$ that has such a limited spatial extent
that ${\sup\Mag{\x'-\x_0}\ll\inf\Mag{\xx'}}$, and the integration
surface $S$, centered on $\x_0$, has a large enough radius
${\Mag{\xx_0}\gg\sup{\Mag{\x'-\x_0}}}$.  Then one can make the usual
far-zone approximation
\begin{align}
\label{eq:approx}
\frac{e^{\im k\Mag{\xx'}}}{\Mag{\xx'}} \approx
\frac{e^{\im k\Mag{\x-\x_0}-\im\k\bdot(\x'-\x_0)}}{\Mag{\xx_0}}
\end{align}
The corresponding approximate retarded time is 
\begin{align}
\tret' =  t' - \frac{\Mag{\xx_0}}{c}
\end{align}
where
\begin{align}
t' \approx t + \frac{\kunit\bdot(\x'-\x_0)}{c}
\end{align}
so that \eqn{eq:E_Thide} for the electric field and
\eqn{eq:B_Jefimenko} for the magnetic field can, in complex notation,
be approximated by
\begin{subequations}
\label{eq:EB_approx}
\begin{align}
\label{eq:E_approx}
\E(t,\x) \approx{}&
 \frac{1}{4\pi\epsz}\,\frac{e^{\im k\Mag{\xx_0}}}{\Mag{\xx_0}^2}\,
 \intV{'}{\rho(t',\x')\kunit} 
\notag\\
 &+\frac{1}{4\pi\epsz c} \,\frac{e^{\im k\Mag{\xx_0}}}{\Mag{\xx_0}^2}\,
 \intV{'}{[\j(t',\x')}\bdot\kunit]\kunit 
\notag\\
 &+\frac{1}{4\pi\epsz c} \,\frac{e^{\im k\Mag{\xx_0}}}{\Mag{\xx_0}^2}\,
 \intV{'}{[\j(t',\x')}\cross\kunit]\cross\kunit 
\notag\\
 &+\frac{1}{4\pi\epsz c^2} \,\frac{e^{\im k\Mag{\xx_0}}}{\Mag{\xx_0}}\,
 \intV{'}{[\jdot(t',\x')}\cross\kunit]\cross\kunit
\intertext{and}
\label{eq:B_approx}
\B(t,\x) \approx{}&
 \frac{1}{4\pi\epsz c^2}\,\frac{e^{\im k\Mag{\xx_0}}}{\Mag{\xx_0}^2}
 \intV{'}{\j(t',\x')}\cross\kunit 
\notag\\
 &+\frac{1}{4\pi\epsz c^3}\,\frac{e^{\im k\Mag{\xx_0}}}{\Mag{\xx_0}}
 \intV{'}{\jdot(t',\x')\cross\kunit}
\end{align}
\end{subequations}
as obtained from inverse Fourier transforming of \eqns{eq:EB_w} with
\eqn{eq:approx} inserted. 

In many cases, when one wants to calculate electromagnetic observables
in the far zone, the approximate \eqns{eq:EB_approx} are accurate
enough. At the same time they are easier to evaluate, and give
results which are physically more lucid than the SPPJ equations
(\ref{eq:EB_Jefimenko}). Below, we follow this scheme to derive general,
untruncated expressions for the linear and angular momentum, valid far
away from the source volume $V'$.

\section{Linear momentum}

The linear momentum conservation law\cite{Schwinger&al:Book:1998}
\begin{align}
\ddt{\pmech} + \ddt{\pfield} + \ointSvec{}{\ten{T}} = \0
\end{align}
describes the balance between the time rate of change of the mechanical
linear momentum $\pmech$, \ie, mechanical force, the time rate of change
of the electromagnetic field linear momentum $\pfield$, and the flow of
linear momentum across the surface $S$ described by the linear momentum
flux tensor density $\ten{T}$ (the negative of Maxwell's stress tensor)
of the electromagnetic field. For example, this law describes how the
translational motion of charges (\eg, a transmitting antenna current)
at one point in space induces a translational motion, via force action,
of remote charges (\eg, a receiving antenna current). This is one of
several physical mechanisms by which information can be transferred
electromagnetically through free space and the primary physical basis
for current radio astronomy and other radio-based science, as well as
today's wireless communications technology.

The linear momentum of the electromagnetic field is defined
as\cite{Panofsky&Phillips:Book:1962,Schwinger&al:Book:1998}
\begin{align}
\label{eq:pfield}
\pfield = \intV{'}{\gfield} 
\end{align}
where
\begin{align}
\gfield = \epsz(\E\cross\B) = \frac{\S}{c^2}
\end{align}
is the electromagnetic field linear momentum density and $\S$ the
Poynting vector. 

Applying the paraxial approximation\cite{Heald&Marion:Book:1995}
\begin{align}
\label{eq:paraxial}
\kunit=\frac{\xx'}{\Mag{\xx'}} \approx \frac{\xx_0}{\Mag{\xx_0}} \equiv \nunit
\end{align}
of the wave vector direction, and moving the constant unit
vector $\nunit$, from $\x_0$ to the
observation point $\x$, outside each of the six constituent integrals in
in the following way
\begin{subequations}
\label{eq:integrals}
\begin{align}
\intV{'}{\rho(t',\x')\kunit} \approx
\Big(\intV{'}{\rho(t',\x')}\Big)\nunit = q(t')\nunit
\end{align}
\begin{align}
\intV{'}{\j(t',\x')}\cross\kunit \approx
\Big(\intV{'}{\j(t',\x')}\Big)\cross\nunit = \I(t')\cross\nunit
\end{align}
\begin{align}
\intV{'}{\jdot(t',\x')}\cross\kunit \approx
\Big(\intV{'}{\jdot(t',\x')}\Big)\cross\nunit = \Idot(t')\cross\nunit
\end{align}
\begin{align}
 \intV{'}{[\j(t',\x')}\bdot\kunit]\kunit &\approx
 \Big[\Big(\intV{'}{\j(t',\x')}\Big)\bdot\nunit\Big]\nunit
\notag\\
 &= (\I\bdot\nunit)\nunit = I_n(t')\nunit
\end{align}
\begin{align}
 \intV{'}{[\j(t',\x')}\cross\kunit]\cross\kunit &\approx
 \Big[\Big(\intV{'}{\j(t',\x')}\Big)\cross\nunit\Big]\cross\nunit
\notag\\
 &= \left(\I\cross\nunit\right)\cross\nunit = I_n(t')\nunit - \I(t')
\end{align}
\begin{align}
 \intV{'}{[\jdot(t',\x')}\cross\kunit]\cross\kunit &\approx
 \Big[\Big(\intV{'}{\jdot(t',\x')}\Big)\cross\nunit\Big]\cross\nunit
\notag\\
 &= \left(\Dot{\I}\cross\nunit\right)\cross\nunit = \Dot{I}_n(t')\nunit - \Idot(t')
\end{align}
\end{subequations}
one obtains, after some vector algebraic simplifications, the following
general, untruncated expression for the cycle averaged linear momentum
density
\begin{multline}
\label{eq:gfield_approx}
\Ave{\gfield} = \frac{1}{32\pi^2\epsz c^3}
\bigg(
 \frac{\Mag{\dot{I}}^2-\Mag{\dot{I}_n}^2}{c^2\Mag{\xx_0}^2}\nunit
\\
 +\frac{2\Re{\I\bdot\cc{\Idot} -I_n\cc{\Dot{I}_n}}
    -\Re{(cq+I_n)\cc{\Dot{I}_n}}} {c\Mag{\xx_0}^3}\nunit
\\
 +\frac{\Mag{I}^2-\Mag{I_n}^2-\Re{(cq+I_n)\cc{I_n}}}{\Mag{\xx_0}^4}\nunit
\\
 +\frac{\Re{(cq+I_n)\cc{\Idot}}}{c\Mag{\xx_0}^3}
 +\frac{\Re{(cq+I_n)\cc{\I}}}{\Mag{\xx_0}^4}
 \bigg)
\end{multline}
This approximate expression is valid far away from the source volume
$V'$.

At very large distances $r\equiv\Mag{\xx_0}$ from the source volume
$V'$, we see that the linear momentum density, and consequently also the
Poynting vector, is accurately represented by the first term, which is
radial (along $\nunit$) and falls off as $1/r^2$, while the successive
non-radial terms fall off as $1/r^3$ or as $1/r^4$, respectively.
However, at finite distances the Poynting vector is \emph{not} radial
but has transverse components. Of course, these transverse components,
which make the Poynting vector spiral around $\nunit$, are difficult to
observe because of their smallness. Furthermore, since the non-radial
terms fall off faster than $1/r^2$, the spiraling of the Poynting vector
will diminish with distance from the source.

\section{Angular momentum}

The angular momentum conservation law\cite{Schwinger&al:Book:1998}
\begin{align}
\ddt{\Jmech(\x_0)} + \ddt{\Jfield(\x_0)} + \ointSvec{'}{\ten{K}(\x_0)} = \0
\end{align}
describes the balance between the time rate of change of the mechanical
angular momentum $\Jmech$, \ie, mechanical torque, the time rate of
change of the electromagnetic field angular momentum $\Jfield$, and
the flow of angular momentum across the surface $S$ described by the
angular momentum flux tensor $\ten{K}=(\xx_0)\cross\ten{T}$, all about
the point $\x_0$. This law describes how the rotational (spin, orbital)
motion of charges at one point in space induces a rotational motion,
via torque action, of charges at other points in space. This is an
additional physical mechanism by which information can be transferred
electromagnetically through free space but one that is used only
partially and sparingly in today's radio-based research and wireless
communication technology.

The field angular momentum about a point $\x_0$ is defined
as\cite{Panofsky&Phillips:Book:1962,Mandel&Wolf:Book:1995,Schwinger&al:Book:1998}
\begin{align}
\label{eq:Jfield}
\Jfield(\x_0) ={}& \intV{'}\hfield(x_0)
\end{align}
where
\begin{align}
\label{eq:hfield}
\hfield(\x_0) = (\xx_0)\cross\gfield
\end{align}
is the electromagnetic angular momentum density.
For a single temporal Fourier component in
complex notation and a beam geometry, \eqn{eq:Jfield} can be
written\cite{Rohrlich:Book:2007,vanEnk&Nienhuis:OC:1992}
\begin{align}
\Jfield = \Lfield + \Sfield 
\end{align}
where
\begin{subequations}
\begin{align}
\Lfield &= -\im\frac{\epsz}{2\w} \intV{'}{\cc{E_i}[(\x'-\x_0)\cross\grad]E_i} \\
\Sfield &= -\im\frac{\epsz}{2\w} \intV{'}(\cc{\E}\cross\E) 
\end{align}
\end{subequations}
The first term is the electromagnetic orbital angular momentum (OAM),
which describes the vorticity of the EM field, and the second term is
the electromagnetic spin angular momentum (SAM), which describes wave
polarization (left or right circular).

Since $\xx_0=\Mag{\xx_0}\nunit$, the vector product in \eqn{eq:hfield}
gives vanishing contributions when operating on the radial terms (terms
parallel to $\nunit$) in \eqn{eq:gfield_approx}. As a result, the
complete cycle averaged far-zone expression for a frequency component
$\w$ of the electromagnetic angular momentum density generated by
arbitrary charge and current sources is simply
\begin{align}
\label{eq:hfield_approx}
\Ave{\hfield(\x_0)} ={}&
\frac{1}{32\pi^2\epsz c^3}
\bigg(
 \frac{\nunit\cross\Re{(cq+I_n)\cc{\Idot}}}{c\Mag{\xx_0}^2}
\notag\\
 &\quad+\frac{\nunit\cross\Re{(cq+I_n)\cc{\I}}}{\Mag{\xx_0}^3}
 \bigg)
\end{align}
We see that at very large distances $r$, the angular
momentum density falls off as $1/r^2$, \ie, it has precisely the same
behavior in the far zone as the linear momentum density and can
therefore also transfer information wirelessly over large distances.
The only difference is that while the direction of the linear momentum
(Poynting vector) becomes purely radial at infinity, the angular momentum
becomes perpendicular to the linear momentum, \ie purely transverse,
there.

\section{Summary and conclusions}

We have shown how the SPPJ equations for the retarded electric
and magnetic fields, generated by arbitrary charge and current
distributions, can be used to derive general far-zone formulas for the
concomitant electromagnetic linear and angular momentum.

From the far-zone approximations obtained for the linear momentum and
angular momentum densities, \eqn{eq:gfield_approx} and \eqn{eq:hfield_approx},
respectively, it is easy to see that both these physical quantities,
to leading order, fall off as $1/r^2$ with distance $r$ from the
source. Consequently, when integrated over a surface element
$r^2\sin(\theta)\upd{\theta}\upd{\varphi}$ of a large spherical shell,
centered on the source, they behave as constants. Physically, this means
that both linear and angular momentum can be carried all the way to
infinity. While the time rate of change of the linear momentum provides
the force that causes the charges in the EM sensor (\eg, a receiving
antenna) perform translational (oscillating) motions, the angular momentum
gives rise to a rotational (spinning and/or orbiting) motion.

However, while the linear momentum is in the far zone determined
essentially by the dominant $\E$ and $\B$ far-zone fields that fall off
as $r^{-1}$, the far-zone angular momentum is \emph{not} determined
by far-zone fields, but rather by \emph{near-zone} components of
the $\E$ field, that fall off as $r^{-2}$, and the far-zone $\B$
field. In fact, the far-zone $\E$ field does not contribute to the
far-zone angular momentum at all! This somewhat surprising result is a
generalization to arbitrary EM fields of a result, originally derived
by \textcite{Abraham:PZ:1914} already in \citeyear{Abraham:PZ:1914},
for the special case of pure dipole fields. To quote from page 916 of
\textcite{Abraham:PZ:1914}(using modern notation):\footnote{In English
translation: ``The electric vector $\E$ has (see Eq.~6c below) a radial
component, which indeed falls off as $r^{-2}$ with increasing $r$, while
the components of $\E$ that are orthogonal to the radius vector fall off
as $r^{-1}$. Hence, with increasing distance from the light source the
light waves become transverse, and the linear momentum becomes parallel
to the radius vector; it could therefore according to (4) seem that the
angular momentum in the wave zone would be equal to zero. However, one
observes that this is not the case, if one determines the order of the
quantities in question; while the longitudinal component of $\E$ is of
the order $r^{-2}$, the scalar product of $\vec{r}$ and $\E$ is still of
order $r^{-1}$, just as $\vec{H}$. From (4c) it therefore follows that
the density of the angular momentum, as well as the densities of the
linear momentum and energy are of the order $r^{-2}$~''}

\begin{quote}
``Der elektrische Vektor $\E$ hat (s.\ Gl.~6c unten) eine radiale
Komponente, die freilich mit wachsendem $r$ wie $r^{-2}$ abnimmt,
während die zum Fahrstrahl senkrechten Komponenten von $\E$ wie $r^{-1}$
abnehmen. Mit wachsender Entfernung von der Lichtquelle werden somit
die Lichtwellen transversal, der Impuls wird parallel dem Fahrstrahl;
es könnte darum nach (4) scheinen, als ob der Drehimpuls der Wellenzone
gleich null sei. Doch sieht man, daß dem nicht so ist, wenn man die
Ordnung der fraglichen Größen bestimmt; wenn auch die longitudinale
Komponente von $\E$ von der Ordnung $r^{-2}$ ist, so ist doch das
skalare Produkt von $\vec{r}$ und $\E$ von der Ordnung $r^{-1}$,
ebenso wie $\vec{H}$. Aus (4c) folgt demnach, daß die Dichte des
Impulsmomentes, ebenso wie die Dichten des Impulses und der Energie, von
der Ordnung $r^{-2}$ ist.''
\end{quote}

The fact that electromagnetic linear and angular momentum can be
transferred wirelessly over long distances (they are both irreversibly
lost at infinity), means, \eg, that both of them can be used in wireless
communications. Today, only the linear momentum and, in some cases the
two orthogonal states of the spin part of the angular momentum (SAM,
wave polarization), are used in wireless communication protocols,
including such modern concepts as MIMO.\cite{Paulraj&al:Book:2003}
However, wave polarization techniques do not always work reliably
in real-world communication settings. A common problem is the
depolarization of the radio beam, due to reflections and other
interactions with the surroundings that can cause a conversion of
polarization (SAM) into orbital angular momentum (OAM). The SAM is
thereby diminished or lost, while the total angular momentum (SAM
plus OAM) is still conserved. The use of OAM, in addition to SAM
(polarization), could therefore be a remedy for this deficiency.
Besides, and perhaps more importantly, OAM spans a denumerably
infinite Hilbert space,\cite{Molina-Terriza&al:PRL:2002} providing,
in principle, an infinite number of orthogonal basis states that
can be used as the ``letters'' of a higher dimensional information
alphabet.\cite{Franke-Arnold&al:LPR:2008,Barreiro&al:NPH:2008} By
adding OAM encoding to an EM beam, this beam can therefore transfer
more wireless information per unit time and unit frequency than a
radio beam carrying only linear momentum (power), or linear momentum
plus SAM (polarization) as is the case for current communication
techniques.\cite{Harwit:APJ:2003,Molina-Terriza&al:NPH:2007}

The possibility to use OAM encoding of EM beams for efficient free-space
information transfer has been successfully demonstrated in several
experiments,\cite{Gibson&al:OE:2004,Lin&al:AO:2007,Wu&Li:CP:2007}
even at the single-photon level.\cite{Mair&al:N:2001} A scheme for
the use of OAM at radio frequencies in the gigahertz range and
below, where digital radio methods are readily applicable to make
the method feasible for practical technology applications, has been
developed.\cite{Thide&al:PRL:2007}

\begin{acknowledgments}
We thank Hans-Jürgen Zepernick and Roger Karlsson for helpful comments.
One of the authors (B.\,T.), gratefully acknowledges the financial
support from the Swedish Research Council (VR).
\end{acknowledgments}


\begin{thebibliography}{24}
\expandafter\ifx\csname natexlab\endcsname\relax\def\natexlab#1{#1}\fi
\expandafter\ifx\csname bibnamefont\endcsname\relax
  \def\bibnamefont#1{#1}\fi
\expandafter\ifx\csname bibfnamefont\endcsname\relax
  \def\bibfnamefont#1{#1}\fi
\expandafter\ifx\csname citenamefont\endcsname\relax
  \def\citenamefont#1{#1}\fi
\expandafter\ifx\csname url\endcsname\relax
  \def\url#1{\texttt{#1}}\fi
\expandafter\ifx\csname urlprefix\endcsname\relax\def\urlprefix{URL }\fi
\providecommand{\bibinfo}[2]{#2}
\providecommand{\eprint}[2][]{\url{#2}}

\bibitem[{\citenamefont{de~Melo~e Souza et~al.}(2009)\citenamefont{de~Melo~e
  Souza, Cougo-Pinto, Farina, and Moriconi}}]{Souza&al:AJP:2009}
\bibinfo{author}{\bibfnamefont{R.}~\bibnamefont{de~Melo~e Souza}},
  \bibinfo{author}{\bibfnamefont{M.~V.} \bibnamefont{Cougo-Pinto}},
  \bibinfo{author}{\bibfnamefont{C.}~\bibnamefont{Farina}}, \bibnamefont{and}
  \bibinfo{author}{\bibfnamefont{M.}~\bibnamefont{Moriconi}},
  \bibinfo{journal}{Am.\ J.~Phys.} \textbf{\bibinfo{volume}{77}},
  \bibinfo{pages}{67} (\bibinfo{year}{2009}).

\bibitem[{\citenamefont{Stratton}(1941)}]{Stratton:Book:1941}
\bibinfo{author}{\bibfnamefont{J.~A.} \bibnamefont{Stratton}},
  \emph{\bibinfo{title}{Electromagnetic Theory}}
  (\bibinfo{publisher}{McGraw-Hill Book Co.}, \bibinfo{address}{New York, NY,
  USA}, \bibinfo{year}{1941}).

\bibitem[{\citenamefont{Panofsky and
  Phillips}(1962)}]{Panofsky&Phillips:Book:1962}
\bibinfo{author}{\bibfnamefont{W.~K.~H.} \bibnamefont{Panofsky}}
  \bibnamefont{and} \bibinfo{author}{\bibfnamefont{M.}~\bibnamefont{Phillips}},
  \emph{\bibinfo{title}{Classical Electricity and Magnetism}}
  (\bibinfo{publisher}{Addison-Wesley Publishing Company},
  \bibinfo{address}{Reading, MA, USA}, \bibinfo{year}{1962}),
  \bibinfo{edition}{2nd} ed., \bibinfo{note}{{ISBN}~0-201-05702-6}.

\bibitem[{\citenamefont{Jefimenko}(1966)}]{Jefimenko:Book:1966}
\bibinfo{author}{\bibfnamefont{O.~D.} \bibnamefont{Jefimenko}},
  \emph{\bibinfo{title}{Electricity and Magnetism. An introduction to the
  theory of electric and magnetic fields}}
  (\bibinfo{publisher}{Appleton-Century-Crofts}, \bibinfo{address}{New York,
  NY, USA}, \bibinfo{year}{1966}).

\bibitem[{\citenamefont{Heald and Marion}(1995)}]{Heald&Marion:Book:1995}
\bibinfo{author}{\bibfnamefont{M.~A.} \bibnamefont{Heald}} \bibnamefont{and}
  \bibinfo{author}{\bibfnamefont{J.~B.} \bibnamefont{Marion}},
  \emph{\bibinfo{title}{Classical Electromagnetic Radiation}}
  (\bibinfo{publisher}{Saunders College Publishing}, \bibinfo{address}{Fort
  Worth, TX, USA}, \bibinfo{year}{1995}), \bibinfo{edition}{3rd} ed.,
  \bibinfo{note}{{ISBN}~0-03-097277-9}.

\bibitem[{\citenamefont{Jackson}(1999)}]{Jackson:Book:1999}
\bibinfo{author}{\bibfnamefont{J.~D.} \bibnamefont{Jackson}},
  \emph{\bibinfo{title}{Classical Electrodynamics}} (\bibinfo{publisher}{Wiley
  \& Sons}, \bibinfo{address}{New York, NY, USA}, \bibinfo{year}{1999}),
  \bibinfo{edition}{3rd} ed., \bibinfo{note}{{ISBN}~0-471-30932-X}.

\bibitem[{Not({\natexlab{a}})}]{Note1}
\bibinfo{note}{We do not consider advanced time solutions.}

\bibitem[{\citenamefont{Schwinger et~al.}(1998)\citenamefont{Schwinger, DeRaad,
  Milton, and Tsai}}]{Schwinger&al:Book:1998}
\bibinfo{author}{\bibfnamefont{J.}~\bibnamefont{Schwinger}},
  \bibinfo{author}{\bibfnamefont{L.~L.} \bibnamefont{DeRaad},
  \bibfnamefont{Jr.}}, \bibinfo{author}{\bibfnamefont{K.~A.}
  \bibnamefont{Milton}}, \bibnamefont{and}
  \bibinfo{author}{\bibfnamefont{W.}~\bibnamefont{Tsai}},
  \emph{\bibinfo{title}{Classical Electrodynamics}}
  (\bibinfo{publisher}{Perseus Books}, \bibinfo{address}{Reading, MA, USA},
  \bibinfo{year}{1998}), \bibinfo{note}{{ISBN}~0-7382-0056-5}.

\bibitem[{\citenamefont{Mendel and Wolf}(1995)}]{Mandel&Wolf:Book:1995}
\bibinfo{author}{\bibfnamefont{L.}~\bibnamefont{Mendel}} \bibnamefont{and}
  \bibinfo{author}{\bibfnamefont{E.}~\bibnamefont{Wolf}},
  \emph{\bibinfo{title}{Optical coherence and quantum optics}}
  (\bibinfo{publisher}{Cambridge University Press}, \bibinfo{address}{New York,
  NY, USA}, \bibinfo{year}{1995}), \bibinfo{note}{{ISBN}~0-521-41711-2}.

\bibitem[{\citenamefont{Rohrlich}(2007)}]{Rohrlich:Book:2007}
\bibinfo{author}{\bibfnamefont{F.}~\bibnamefont{Rohrlich}},
  \emph{\bibinfo{title}{Classical Charged Particles}}
  (\bibinfo{publisher}{World Scientific}, \bibinfo{address}{Singapore},
  \bibinfo{year}{2007}), \bibinfo{edition}{3rd} ed.

\bibitem[{\citenamefont{van Enk and Nienhuis}(1992)}]{vanEnk&Nienhuis:OC:1992}
\bibinfo{author}{\bibfnamefont{S.~J.} \bibnamefont{van Enk}} \bibnamefont{and}
  \bibinfo{author}{\bibfnamefont{G.}~\bibnamefont{Nienhuis}},
  \bibinfo{journal}{Opt. Commun.} \textbf{\bibinfo{volume}{94}},
  \bibinfo{pages}{147} (\bibinfo{year}{1992}).

\bibitem[{\citenamefont{Abraham}(1914)}]{Abraham:PZ:1914}
\bibinfo{author}{\bibfnamefont{M.}~\bibnamefont{Abraham}},
  \bibinfo{journal}{Physik. Zeitschr.} \textbf{\bibinfo{volume}{XV}},
  \bibinfo{pages}{914} (\bibinfo{year}{1914}).

\bibitem[{Not({\natexlab{b}})}]{Note2}
\bibinfo{note}{In English translation: ``The electric vector $\protect \mathbf
  {E}$ has (see Eq.~6c below) a radial component, which indeed falls off as
  $r^{-2}$ with increasing $r$, while the components of $\protect \mathbf {E}$
  that are orthogonal to the radius vector fall off as $r^{-1}$. Hence, with
  increasing distance from the light source the light waves become transverse,
  and the linear momentum becomes parallel to the radius vector; it could
  therefore according to (4) seem that the angular momentum in the wave zone
  would be equal to zero. However, one observes that this is not the case, if
  one determines the order of the quantities in question; while the
  longitudinal component of $\protect \mathbf {E}$ is of the order $r^{-2}$,
  the scalar product of $\protect \mathbf {r}$ and $\protect \mathbf {E}$ is
  still of order $r^{-1}$, just as $\protect \mathbf {H}$. From (4c) it
  therefore follows that the density of the angular momentum, as well as the
  densities of the linear momentum and energy are of the order $r^{-2}$~''}.

\bibitem[{\citenamefont{Paulraj et~al.}(2003)\citenamefont{Paulraj, Nabar, and
  Gore}}]{Paulraj&al:Book:2003}
\bibinfo{author}{\bibfnamefont{A.}~\bibnamefont{Paulraj}},
  \bibinfo{author}{\bibfnamefont{R.}~\bibnamefont{Nabar}}, \bibnamefont{and}
  \bibinfo{author}{\bibfnamefont{D.}~\bibnamefont{Gore}},
  \emph{\bibinfo{title}{Introduction to Space-Time Wireless Communications}}
  (\bibinfo{publisher}{Cambridge Univ. Press}, \bibinfo{address}{Cambridge,
  UK}, \bibinfo{year}{2003}), \bibinfo{note}{{ISBN}~0-521-82615-2}.

\bibitem[{\citenamefont{Molina-Terriza
  et~al.}(2002)\citenamefont{Molina-Terriza, Torres, and
  Torner}}]{Molina-Terriza&al:PRL:2002}
\bibinfo{author}{\bibfnamefont{G.}~\bibnamefont{Molina-Terriza}},
  \bibinfo{author}{\bibfnamefont{J.~P.} \bibnamefont{Torres}},
  \bibnamefont{and} \bibinfo{author}{\bibfnamefont{L.}~\bibnamefont{Torner}},
  \bibinfo{journal}{Phys.\ Rev.\ Lett.} \textbf{\bibinfo{volume}{3}},
  \bibinfo{pages}{013601,4 pp} (\bibinfo{year}{2002}).

\bibitem[{\citenamefont{Franke-Arnold et~al.}(2008)\citenamefont{Franke-Arnold,
  Allen, and Padgett}}]{Franke-Arnold&al:LPR:2008}
\bibinfo{author}{\bibfnamefont{S.}~\bibnamefont{Franke-Arnold}},
  \bibinfo{author}{\bibfnamefont{L.}~\bibnamefont{Allen}}, \bibnamefont{and}
  \bibinfo{author}{\bibfnamefont{M.}~\bibnamefont{Padgett}},
  \bibinfo{journal}{Laser \& Photon.\ Rev.} \textbf{\bibinfo{volume}{2}},
  \bibinfo{pages}{299} (\bibinfo{year}{2008}).

\bibitem[{\citenamefont{Barreiro et~al.}(2008)\citenamefont{Barreiro, Wei, and
  Kwiat}}]{Barreiro&al:NPH:2008}
\bibinfo{author}{\bibfnamefont{J.~T.} \bibnamefont{Barreiro}},
  \bibinfo{author}{\bibfnamefont{T.-C.} \bibnamefont{Wei}}, \bibnamefont{and}
  \bibinfo{author}{\bibfnamefont{P.~W.} \bibnamefont{Kwiat}},
  \bibinfo{journal}{Nature Phys.} \textbf{\bibinfo{volume}{4}},
  \bibinfo{pages}{282} (\bibinfo{year}{2008}).

\bibitem[{\citenamefont{Harwit}(2003)}]{Harwit:APJ:2003}
\bibinfo{author}{\bibfnamefont{M.}~\bibnamefont{Harwit}},
  \bibinfo{journal}{Astrophys.\ J.} \textbf{\bibinfo{volume}{597}},
  \bibinfo{pages}{1266} (\bibinfo{year}{2003}).

\bibitem[{\citenamefont{Molina-Terriza
  et~al.}(2007)\citenamefont{Molina-Terriza, Torres, and
  Torner}}]{Molina-Terriza&al:NPH:2007}
\bibinfo{author}{\bibfnamefont{G.}~\bibnamefont{Molina-Terriza}},
  \bibinfo{author}{\bibfnamefont{J.~P.} \bibnamefont{Torres}},
  \bibnamefont{and} \bibinfo{author}{\bibfnamefont{L.}~\bibnamefont{Torner}},
  \bibinfo{journal}{Nature Phys.} \textbf{\bibinfo{volume}{3}},
  \bibinfo{pages}{305} (\bibinfo{year}{2007}).

\bibitem[{\citenamefont{Gibson et~al.}(2004)\citenamefont{Gibson, Courtial,
  Padgett, Vasnetsov, Pas'ko, Barnett, and Franke-Arnold}}]{Gibson&al:OE:2004}
\bibinfo{author}{\bibfnamefont{G.}~\bibnamefont{Gibson}},
  \bibinfo{author}{\bibfnamefont{J.}~\bibnamefont{Courtial}},
  \bibinfo{author}{\bibfnamefont{M.~J.} \bibnamefont{Padgett}},
  \bibinfo{author}{\bibfnamefont{M.}~\bibnamefont{Vasnetsov}},
  \bibinfo{author}{\bibfnamefont{V.}~\bibnamefont{Pas'ko}},
  \bibinfo{author}{\bibfnamefont{S.~M.} \bibnamefont{Barnett}},
  \bibnamefont{and}
  \bibinfo{author}{\bibfnamefont{S.}~\bibnamefont{Franke-Arnold}},
  \bibinfo{journal}{Opt. Express} \textbf{\bibinfo{volume}{12}},
  \bibinfo{pages}{5448} (\bibinfo{year}{2004}).

\bibitem[{\citenamefont{Lin et~al.}(2007)\citenamefont{Lin, Yuan, Tao, and
  Burge}}]{Lin&al:AO:2007}
\bibinfo{author}{\bibfnamefont{J.}~\bibnamefont{Lin}},
  \bibinfo{author}{\bibfnamefont{X.-C.} \bibnamefont{Yuan}},
  \bibinfo{author}{\bibfnamefont{S.~H.} \bibnamefont{Tao}}, \bibnamefont{and}
  \bibinfo{author}{\bibfnamefont{R.~E.} \bibnamefont{Burge}},
  \bibinfo{journal}{Appl.\ Opt.} \textbf{\bibinfo{volume}{46}},
  \bibinfo{pages}{4680} (\bibinfo{year}{2007}).

\bibitem[{\citenamefont{Wu and Li}(2007)}]{Wu&Li:CP:2007}
\bibinfo{author}{\bibfnamefont{J.-Z.} \bibnamefont{Wu}} \bibnamefont{and}
  \bibinfo{author}{\bibfnamefont{Y.-J.} \bibnamefont{Li}},
  \bibinfo{journal}{Chin.\ Phys.} \textbf{\bibinfo{volume}{16}},
  \bibinfo{pages}{1334} (\bibinfo{year}{2007}).

\bibitem[{\citenamefont{Mair et~al.}(2001)\citenamefont{Mair, Vaziri, Weihs,
  and Zeilinger}}]{Mair&al:N:2001}
\bibinfo{author}{\bibfnamefont{A.}~\bibnamefont{Mair}},
  \bibinfo{author}{\bibfnamefont{A.}~\bibnamefont{Vaziri}},
  \bibinfo{author}{\bibfnamefont{G.}~\bibnamefont{Weihs}}, \bibnamefont{and}
  \bibinfo{author}{\bibfnamefont{A.}~\bibnamefont{Zeilinger}},
  \bibinfo{journal}{Nature} \textbf{\bibinfo{volume}{412}},
  \bibinfo{pages}{313} (\bibinfo{year}{2001}).

\bibitem[{\citenamefont{Thid{\'e} et~al.}(2007)\citenamefont{Thid{\'e}, Then,
  Sj{\"o}holm, Palmer, Bergman, Carozzi, Istomin, Ibragimov, and
  Khamitova}}]{Thide&al:PRL:2007}
\bibinfo{author}{\bibfnamefont{B.}~\bibnamefont{Thid{\'e}}},
  \bibinfo{author}{\bibfnamefont{H.}~\bibnamefont{Then}},
  \bibinfo{author}{\bibfnamefont{J.}~\bibnamefont{Sj{\"o}holm}},
  \bibinfo{author}{\bibfnamefont{K.}~\bibnamefont{Palmer}},
  \bibinfo{author}{\bibfnamefont{J.}~\bibnamefont{Bergman}},
  \bibinfo{author}{\bibfnamefont{T.~D.} \bibnamefont{Carozzi}},
  \bibinfo{author}{\bibfnamefont{Y.~N.} \bibnamefont{Istomin}},
  \bibinfo{author}{\bibfnamefont{N.~H.} \bibnamefont{Ibragimov}},
  \bibnamefont{and}
  \bibinfo{author}{\bibfnamefont{R.}~\bibnamefont{Khamitova}},
  \bibinfo{journal}{Phys.\ Rev.\ Lett.} \textbf{\bibinfo{volume}{99}},
  \bibinfo{pages}{087701(4)} (\bibinfo{year}{2007}).

\end{thebibliography}

\end{document}